

\mathchardef\alpha="710B \mathchardef\beta="710C
\mathchardef\gamma="710D \mathchardef\delta="710E
\mathchardef\epsilon="710F \mathchardef\zeta="7110
\mathchardef\eta="7111 \mathchardef\theta="7112 \mathchardef\iota="7113
\mathchardef\kappa="7114 \mathchardef\lambda="7115
\mathchardef\mu="7116 \mathchardef\nu="7117 \mathchardef\xi="7118
\mathchardef\pi="7119 \mathchardef\rho="711A \mathchardef\sigma="711B
\mathchardef\tau="711C \mathchardef\upsilon="711D
\mathchardef\phi="711E \mathchardef\chi="711F \mathchardef\psi="7120
\mathchardef\omega="7121 \mathchardef\varepsilon="7122
\mathchardef\vartheta="7123 \mathchardef\varpi="7124
\mathchardef\varrho="7125 \mathchardef\varsigma="7126
\mathchardef\varphi="7127 \mathchardef\Gamma="7000
\mathchardef\Delta="7001 \mathchardef\Theta="7002
\mathchardef\Lambda="7003 \mathchardef\Xi="7004 \mathchardef\Pi="7005
\mathchardef\Sigma="7006 \mathchardef\Upsilon="7007
\mathchardef\Phi="7008 \mathchardef\Psi="7009 \mathchardef\Omega="700A
\mathchardef\aleph="7240 \def\hbar{\mathchar'26\mkern-9mu h}
\def\bhbar{{\bf\mathchar'26}\mkern-9mu\bmit h}

\mathchardef\imath="717B \mathchardef\jmath="717C
\mathchardef\ell="7160 \mathchardef\wp="717D \mathchardef\Re="723C
\mathchardef\Im="723D \mathchardef\partial="7140
\mathchardef\infty="7231 \mathchardef\prime="7230
\mathchardef\emptyset="723B \mathchardef\nabla="7272
\mathchardef\dalem="0A32 \mathchardef\top="723E \mathchardef\bot="723F
\mathchardef\triangle="7234 \mathchardef\forall="7238
\mathchardef\exists="7239 \mathchardef\neg="723A 
\mathchardef\flat="715B \mathchardef\natural="715C
\mathchardef\sharp="715D \mathchardef\clubsuit="727C
\mathchardef\diamondsuit="727D \mathchardef\heartsuit="727E
\mathchardef\spadesuit="727F \mathchardef\triangleleft="712F
\mathchardef\triangleright="712E \mathchardef\bigtriangleup="7234
\mathchardef\bigtriangledown="7235

\newfam\itfam \newfam\slfam \newfam\bffam \newfam\ttfam \newfam\bmitfam
\newfam\bcalfam \newfam\lafam



\font\fiverm=cmr5 \font\fivemit=cmmi5 \font\fivecal=cmsy5
\font\fivebf=cmbx5


\font\sevenrm=cmr7 \font\sevenit=cmti7 \font\sevenmit=cmmi7
\font\sevencal=cmsy7 \font\sevenbf=cmbx7


\font\ninerm=cmr9 \font\nineit=cmti9 \font\ninemit=cmmi9
\font\ninecal=cmsy9 \font\ninebf=cmbx9


\font\tenrm=cmr10 \font\tenmit=cmmi10 \font\tenbmit=cmmi10
\font\tenex=cmex10 \font\tenit=cmti10 \font\tencal=cmsy10
\font\tensl=cmsl10 \font\tenbf=cmbx10 \font\tentt=cmtt10
\font\tenbcal=cmbsy10 \font\tenlasy=lasy10


\font\twelverm=cmr10 scaled \magstep1
\font\twelvemit=cmmi10 scaled \magstep1
\font\twelvebmit=cmmi10 scaled \magstep1
\font\twelvecal=cmsy10 scaled \magstep1
\font\twelvebcal=cmbsy10 scaled \magstep1
\font\twelveex=cmex10 scaled \magstep1
\font\twelveit=cmti10 scaled \magstep1
\font\twelvesl=cmsl10 scaled \magstep1
\font\twelvebf=cmbx10 scaled \magstep1
\font\twelvett=cmtt10 scaled \magstep1
\font\twelvelasy=lasy10 scaled \magstep1


\font\fourteenrm=cmr10 scaled \magstep2
\font\fourteenmit=cmmi10 scaled \magstep2
\font\fourteenbmit=cmmi10 scaled \magstep2
\font\fourteenit=cmti10 scaled \magstep2
\font\fourteencal=cmsy10 scaled \magstep2
\font\fourteensl=cmsl10 scaled \magstep2
\font\fourteenbf=cmbx10 scaled \magstep2
\font\fourteentt=cmtt10 scaled \magstep2
\font\fourteenex=cmex10 scaled \magstep2
\font\fourteenbcal=cmbsy10 scaled \magstep2

\message{main font groups,}

\catcode`@=11
\newdimen\z@
\z@=0pt
\newskip\ttglue


\def\ninepoint{\def\rm{\fam0\ninerm}
\abovedisplayskip 11pt plus 2pt minus 8pt
\belowdisplayskip 11pt plus 2pt minus 8pt
\abovedisplayshortskip 0pt plus 2pt
\belowdisplayshortskip 7pt plus 2pt minus 3pt
\textfont0=\ninerm \scriptfont0=\sevenrm \scriptscriptfont0=\fiverm
\textfont1=\ninemit  \scriptfont1=\sevenmit  \scriptscriptfont1=\fivemit
\textfont2=\ninecal \scriptfont2=\sevencal \scriptscriptfont2=\fivecal
\textfont\itfam=\nineit \scriptfont\itfam=\sevenit
\def\cal{\fam2} \def\it{\fam\itfam\nineit} \def\mit{\fam1}
\def\bf{\fam\bffam\ninebf} \normalbaselineskip=20pt
\setbox\strutbox=\hbox{\vrule height7.5pt depth2.5pt width\z@}%
\let\sc=\ninerm \normalbaselines   \rm}


\def\tenpoint{\def\rm{\fam0\tenrm}

\abovedisplayskip 12pt plus 3pt minus 9pt
\belowdisplayskip 12pt plus 3pt minus 9pt
\abovedisplayshortskip 0pt plus 3pt
\belowdisplayshortskip 7pt plus 3pt minus 4pt
\textfont0=\tenrm \scriptfont0=\sevenrm \scriptscriptfont0=\fiverm
\textfont1=\tenmit  \scriptfont1=\sevenmit  \scriptscriptfont1=\fivemit
\textfont2=\tencal \scriptfont2=\sevencal \scriptscriptfont2=\fivecal
\textfont3=\tenex \scriptfont3=\tenex  \scriptscriptfont3=\tenex
\textfont\itfam=\tenit \textfont\slfam=\tensl \textfont\bffam=\tenbf
\scriptfont\bffam=\sevenbf \scriptscriptfont\bffam=\fivebf
\textfont\ttfam=\tentt \textfont\bmitfam=\tenbmit
\textfont\bcalfam=\tenbcal \textfont\lafam=\tenlasy
\def\bmit{\fam\bmitfam\tenbmit} \def\bcal{\fam\bcalfam\tenbcal}
\def\cal{\fam2} \def\it{\fam\itfam\tenit} \def\sl{\fam\slfam\tensl}
\def\mit{\fam1} \def\bf{\fam\bffam\tenbf} \def\ex{\fam3}
\def\la{\fam\lafam\tenlasy} \def\tt{\fam\ttfam\tentt} \tt \ttglue=.5em
plus.25em minus.15em \normalbaselineskip=22pt
\setbox\strutbox=\hbox{\vrule height8.5pt depth3.5pt width\z@}%
\let\sc=\ninerm \normalbaselines   \rm}


\def\twelvepoint{\def\rm{\fam0\twelverm}%
  \abovedisplayskip 14pt plus 3.6pt minus 10.8pt
  \belowdisplayskip 14pt plus 3.6pt minus 10.8pt
  \abovedisplayshortskip 0pt plus 3.6pt
  \belowdisplayshortskip 8.4pt plus 3.6pt minus 4.8pt
     \textfont0=\twelverm \scriptfont0=\ninerm \scriptscriptfont0=\sevenrm%
\textfont1=\twelvemit \scriptfont1=\ninemit \scriptscriptfont1=\sevenmit%
\textfont2=\twelvecal \scriptfont2=\ninecal \scriptscriptfont2=\sevencal%
\textfont3=\twelveex \scriptfont3=\twelveex \scriptscriptfont3=\twelveex%
\def\mit{\fam1\twelvemit}
\def\cal{\fam2\twelvecal}
\def\ex{\fam3\twelveex}
\textfont\bmitfam=\twelvebmit\scriptfont\bmitfam=\ninemit
\def\bmit{\fam\bmitfam\twelvebmit}%
\textfont\bcalfam=\twelvebcal  \def\bcal{\fam\bcalfam\twelvebcal}%
\textfont\itfam=\twelveit \def\it{\fam\itfam\twelveit}%
\textfont\slfam=\twelvesl \def\sl{\fam\slfam\twelvesl}%
\textfont\bffam=\twelvebf \scriptfont\bffam=\ninebf
\scriptscriptfont\bffam=\sevenbf \def\bf{\fam\bffam\twelvebf}%
\textfont\ttfam=\twelvett \def\tt{\fam\ttfam\twelvett}%
\tt \ttglue=.5em plus.25em minus.15em
\textfont\lafam\twelvelasy
\def\la{\lafam\twelvelasy}
\normalbaselineskip=26pt
\setbox\strutbox=\hbox{\vrule height10.5pt depth5.0pt width\z@}
\let\sc=\tenrm  \normalbaselines\rm}


\def\fourteenpoint{\def\rm{\fam0\fourteenrm}
\abovedisplayskip 16pt plus 4.2pt minus 12.6pt
\belowdisplayskip 16pt plus 4.2pt minus 12.6pt
\abovedisplayshortskip 0pt plus 4.2pt
\belowdisplayshortskip 9.8pt plus 4.2pt minus 5.4pt
\textfont0=\fourteenrm \scriptfont0=\tenrm  \scriptscriptfont0=\ninerm
\textfont1=\fourteenmit  \scriptfont1=\tenmit   \scriptscriptfont1=\ninemit
\textfont2=\fourteencal \scriptfont2=\tencal  \scriptscriptfont2=\ninecal
\textfont3=\fourteenex \scriptfont3=\fourteenex \scriptscriptfont3=\fourteenex
\textfont\itfam=\fourteenit \textfont\slfam=\fourteensl
\textfont\bffam=\fourteenbf \scriptfont\bffam=\tenbf
\scriptscriptfont\bffam=\ninebf \textfont\ttfam=\fourteentt
\textfont\bmitfam=\fourteenbmit \scriptfont\bmitfam=\tenbmit
\textfont\bcalfam=\fourteenbcal \scriptfont\bcalfam=\tenbcal
\def\bmit{\fam\bmitfam\fourteenbmit}
\def\bcal{\fam\bcalfam\fourteenbcal} \def\cal{\fam2}
\def\it{\fam\itfam\fourteenit} \def\sl{\fam\slfam\fourteensl}
\def\mit{\fam1} \def\bf{\fam\bffam\fourteenbf} \def\ex{\fam3}
\def\tt{\fam\ttfam\fourteentt}
\tt \ttglue=.5em plus.25em minus.15em
\normalbaselineskip=30pt
\setbox\strutbox=\hbox{\vrule height12.5pt depth6.5pt width\z@}%
\let\sc=\twelverm\normalbaselines\nonfrenchspacing    \rm}
\catcode`@=12

\def\hbar{{\mathchar'26\kern-.5em{\it h}}}
\def\bhbar{{\mathchar'26\kern-.5em{\bmit h}}}

\def\gtorder{\mathrel{\raise.3ex\hbox{$>$}\mkern-14mu
             \lower0.6ex\hbox{$\sim$}}}
\def\ltorder{\mathrel{\raise.3ex\hbox{$<$}\mkern-14mu
             \lower0.6ex\hbox{$\sim$}}}


\catcode`@=11
\def\footnote#1{\edef\@sf{\spacefactor\the\spacefactor}#1\@sf
     \insert\footins\bgroup\tenpoint
     \interlinepenalty100 \let\par=\endgraf
       \leftskip=0pt\rightskip=0pt
       \splittopskip=10pt plus 1pt minus 1pt \floatingpenalty=20000
       \smallskip\item{#1}\bgroup\strut\aftergroup\@foot\let\next}
\catcode`@=12


\def\toppagenumbers{\footline={\hfil
                              }
                    \headline={\ifnum\pageno=1\hfil
                               \else\hss\tenrm\folio\hss \fi
                              }
                   }


      \def\pmb#1{\setbox0=\hbox{#1}%
           \kern-.025em\copy0\kern-\wd0
           \kern.05em\copy0\kern-\wd0
           \kern-.025em\raise.0433em\box0 }


\overfullrule=0pt


\outer\def\newsect#1\par{\smallskip\noindent\line{\bf #1\hfill}\nobreak%
              \smallskip\noindent}

\voffset=0pt
\hoffset=0pt
\hsize=16.5truecm
\vsize=22truecm
\twelvepoint

\belowdisplayskip=24truept
\abovedisplayskip=24truept
\hfill CITA/93/50

\baselineskip=14truept plus 1pt minus 2pt
$  $
\vskip 1truein

\centerline{\fourteenbf  Neutrino Lasing in the Sun}

\bigskip
\bigskip
\centerline{R.A.\ Malaney$^1$, G.D.\ Starkman$^{2,1}$, M.N.\ Butler$^3$}
\vskip 0.5truein

\midinsert\narrower\narrower
\it\parindent=0pt\obeylines

$^1$Canadian Institute for Theoretical Astrophysics, University of Toronto,
\ \ Toronto ON, CANADA M5S 1A7.
$^2$Canadian Institute for Advanced  Research, Cosmology Program.
$^3$Department\ of Physics, Saint Mary's University,  Halifax, NS,
\ \ CANADA B3H 3C3.
\endinsert

\vskip 1truein

\baselineskip=24truept plus 1pt minus 2pt

\midinsert\narrower\narrower

Applying the phenomenon of neutrino lasing in the solar interior, we
show how the  rate for the generic neutrino decay process
$\nu\rightarrow {\rm fermion} +{\rm boson}$, can in principal be
enhanced by many orders of magnitude over its normal decay rate.  Such
a large enhancement could be of import to neutrino-decay models
invoked  in response to the apparent deficit of  electron neutrinos
observed from the sun.  The significance of this result to such models
depends on the specific form of the neutrino decay, and  the particle
model within which it is embedded.

\endinsert

\bigskip
\bigskip
\noindent{\bf PACS number(s):12.15.Cc, 13.35.+s, 14.60.Gh, 96.60.Kx}
\vfil
\eject

The  Solar Neutrino Problem (SNP) is well established. Neutrino fluxes
predicted by standard solar models [1] are in disagreement with the
fluxes observed by current experiments. In terms of the ratio of
observed-to-predicted neutrino fluxes, $\eta$, the Cl-Ar experiment [2]
reports $\eta_{Cl}=0.28 \pm 0.03$; whereas the Kamiokande collaboration
[3] report $\eta_K=0.51\pm 0.07$. These two experiments are mostly
sensitive to the high-energy $^8$B neutrinos.  In order to measure the
lower-energy pp neutrinos, two Ga-Ge experiments have been attempted.
The {\it SAGE} collaboration [4] report $\eta_{SAGE}=0.53\pm 0.19$;
whereas, the {\it GALLEX} collaboration [5] report $\eta_{GALL}=0.62\pm
0.15$. The errors are the $1\sigma$ statistical and systematic errors
reported by the different groups added in quadrature: theoretical
uncertainties from the standard solar model are not included.

Neutrino decay is often invoked as a possible remedy to the the above
discrepancies [6,7].  Although $\nu_e$ decay as a solution to the SNP
would appear to be in contradiction with the observed $\tilde\nu_e$
pulse from SN1987A, a viable solution remains where the mass
eigenstates $\nu_1\>,\nu_2$ are a substantial mixture of the weak
eigenstates $\nu_e\>,\nu_x$ ($x=\mu,\tau$) [8].  Matter induced effects
can also require some additional interpretation of the supernova limits
[9].   The generic neutrino decay can  be
described $\nu_2\rightarrow  \nu_1  + B$, where $B$ is some unknown boson.
The most favored identification of the boson is with some type of majoron
particle $\phi$. For example, the two decay modes
$\nu_2\rightarrow  \nu_1  + \phi$ and
$ \nu_2\rightarrow \>\tilde \nu_1 +\phi$, have been  previously
analysed in detail, and regions of the lifetime and mixing-angle
phase-space consistent with current observations determined [7].  Fast
majoron decay into sterile neutrino states has also been discussed with
regard to the SNP [10].

The main purpose of this report is to point out that the phenomenon of
{\it neutrino lasing} may be of import with regard to the  generic
neutrino decay paradigm.  Specifically, we will find that the bosons
emitted in relativistic neutrino decays can stimulate the decay to
proceed orders of magnitude faster than that described by the normal
decay rate.  Regions in the allowed parameter space consistent with the
observed neutrino fluxes are therefore  subject to  revision.

The phenomenon of neutrino lasing in the context of the early universe
and the dark-matter problem has previously been discussed [11].
Neutrino lasing can best be described as a process in which the decay
of a relativistic neutrino proceeds by  stimulated emission of bosons,
thereby dramatically increasing its  decay rate. We wish to see how solar
neutrinos can have their decay rates affected by this phenomenon.
Considering the generic decay (and its inverse) of a heavy neutrino,
$H$, into a lighter neutrino, $F$ and a boson $B$, it can be shown that
[11] the  evolution of the occupation number  distributions $f_B$, is
described by the following Boltzmann equation:
\hfill
\medskip
$$
{\dot f_B} (E_B)=   {m_H^2 \Gamma_0 \over m_0 E_B p_B}
\int^{E_H^+(E_B)}_{E_H^-(E_B)} dE_{H}
[f_H (1 - f_F)(1+ f_B) - f_B f_F(1 - f_H)]  \eqno(1)
$$
where $p_B$ is the boson three-momentum, $\Gamma_0$ is the free
decay rate for a $H$ at rest,
 and $m_0/2$ is the three-momentum of the decay products in the $H$
rest-frame. Here we have assumed isotropic decay in the $H$ rest
frame. In the massless boson limit we have
$$
m_0 = m_H -   {m_F^2\over m_H} \>. \eqno(2)$$
The integration is over the energy-conserving plane $E_H = E_F +E_B$,
with limiting values
$$E_H^+ = \infty\>, \ \ \ \ \ E_H^- = {m_0 m_H \over 2}
\biggl( {2E_B \over m_0^2} + {1\over 2E_B}\biggr)\>,
\eqno(3a)
$$
and
$$E_{B,F}^{\pm} = (m_0 / 2 m_H)(E_H  \pm p_H)\>. \eqno(3b)$$

 Considering first  only the initial growth of $\dot f_B$, our
Boltzmann equation becomes
$$\dot f_B\approx{E_H^2\over m_H^3}f_B\Gamma_0\int dEf_H  \>,\eqno(4)$$
where we have taken the growth rate to be maximized at
$E_B^-  \ (\approx m_H^2/E_H)$.  We can approximate
$$\int dE f_H \sim {n_\nu\over E_H^3}E_H \>.\eqno(5)$$
Therefore,
$$\dot f_B\approx{n_\nu\over E_H^3}({E_H\over m_H})^3\Gamma_0 f_B
\eqno(6)$$
For significant decay of particles we require $p_B^3f_B\sim p_H^3 f_H$,
that is we need
$$f_B\sim ({E_H^3 n_\nu\over m_H^6}) \>. \eqno(7)$$
Therefore from eqs (6) and (7) we find that the time required
for the number of bosons to equal the number of neutrinos at some
point in the sun is given by
$$t\sim \Gamma_0^{-1}\bigg[{E_H^3\over n_\nu}
\bigg({m_H\over E_H}\bigg)^3
\ln\bigg({E_H^3n_\nu\over m_H^6}\bigg)\bigg]
  \>,  \eqno(8)$$
which typically is much less than the age of the sun.

Using $\Gamma_0=(E_H/m_H) \Gamma_\odot$, where $\Gamma_\odot$ is the
decay rate in the solar frame, Eq.~(6) allows us to define a new
effective decay rate in the presence of lasing as
$$\Gamma_{lase} \sim\biggl [{n_\nu\over E_H^3}
{\biggl ({E_H\over m_H}\biggr )^4 \biggr ]   \Gamma_\odot}
  \>.  \eqno(9)$$
As an example of the effect consider the decay of the pp neutrinos as
they pass through the point $R=0.1 R_{\odot}$.  From the  pp neutrino
density at $0.1 R_{\odot}$, we find ${n_\nu\over
E_H^3}\sim 10^{-25}$, and therefore for $E_H/m_H\gtorder 10^{7}$, the
decay rate is enhanced.  For example, if the mass of the electron
neutrino was $m_{H}\sim 0.01$ eV, the lasing rate is a factor $\sim
10^6$ larger than $\Gamma_\odot$.  Clearly for even smaller values of
the neutrino mass the effect on the decay rate will become more
significant.  This would be the case as long as $E_B^- \gtorder 1/R$,
since if this bound is not satisfied the above analysis would break
down as the wavelength of the emitted boson would be larger than the
size of the region producing it. For practical purposes, we can impose
the limit $m_{H}>10^{-14}$ eV as the mass above which our analysis
would be valid.  Another limitation of our analysis is that we have
neglected transport terms in the Boltzmann equation. This is a good
approximation as long as $\Gamma_{lase}>> 1/R$, since then $f_b$ will
be able to grow.  Finally, note  from the $f$ terms of Eq.~(1) that the
lasing effect ``switches off'' when the  abundance of decaying
neutrinos equals that of the decay-product neutrinos.

As mentioned earlier, we have assumed in the above analysis that decays
in the rest frame of the decaying neutrino are isotropic. What form
does Eq.~(9) take if we drop this assumption?  In the more general
case, the Boltzmann equation Eq.~(1) takes the form [12]
$$
\eqalignno{
{\dot f_B} =   {m_H^3 \Gamma_0 \over m_0 E_B p_B}
\int^{E_H^+(E_B)}_{E_H^-(E_B)} dE_{H}
& [f_H (1 - f_F)(1+ f_B) - f_B f_F(1 - f_H)]\cr
&\times  \Biggl[ {1\over m_H} + \alpha{E_B^+ + E_B^- - 2E_B\over
 p_H m_o}\Biggr]&(10)\cr
}
$$
where $\alpha$ is defined through the probability distribution
$P(\theta)$ given by
$$
P(\theta)={1 \over 2} (1-\alpha \cos \theta)  \>,\eqno(11)
$$
and where $\theta$ is the angle in the rest frame between the decay
product velocity and the  parent velocity in the solar frame.  The
isotropic case discussed previously corresponds to $\alpha=0$, and is
valid (up to factors of $\sim 2$) for all $\alpha > -1$.  For $\alpha
\simeq -1$, the emission of low-momentum bosons is suppressed.
Utilizing Eq.~(10), and applying similar arguments as those used in the
isotropic case we find the equivalent equation to Eq.~(9) is
$$\Gamma_{lase} \sim\biggl [{n_\nu\over E_H^3}{\biggl
({E_H\over m_H}\biggr )^2 \biggr ]   \Gamma_\odot}
  \>.  \eqno(12)$$
 From comparison of Eq.~(9) with (12) it can be seen that $\alpha=-1$
results in a suppression factor $(m_H/E_H)^2$, relative to the case of
isotropic decays.

Let us consider the pp neutrinos produced by the sun, and focus on the
decay of these neutrinos into a tau or muon neutrino ($\nu_e
\rightarrow \nu_x + B$).  In order to accurately assess the importance
of neutrino lasing we must average the effect over the varying matter
and neutrino densities in the sun.  Let us assume that matter effects
dominate the neutrino masses. In this case we can replace $m_H$ in
Eqs.~(8) and (12) with ${\sqrt{2EV}}$, where $E$ is the $\nu_e$ energy
and the effective potential $V$ is  given by $$V=\sqrt 2 G_F \rho
m_n^{-1} Y_e\>. \eqno(13)$$ Here, $G_F$ is the Fermi constant, $\rho$
is the matter density, $m_n$ is the nucleon mass, and $Y_{e}$ is the
number of electrons per nucleon.

In the solar frame, the decay rate for some process described by some
dimensionless coupling constant $g$ can be written
$$\Gamma_\odot={g^2 m^2_H\over 16 \pi E} \equiv {g^2 V\over 16 \pi}
\>. \eqno(14)$$
Neglecting lasing effects, in terms of the electron-neutrino flux
$\Phi_e^o$ at some origin, the flux $\Phi_e(r)$ at distance $r$  from
the origin is given by [13]
$$\Phi_e(r)=\Phi_e^o\exp\biggl [-{1\over 8 d_o} \sum_x g^2 d^{eff}
\biggr ] \>,
\eqno(15)$$
where the effective matter width $d^{eff}$  for the decay is given by
$$d^{eff}=\int^r_0 \rho(r) Y_e(r)dr \>,\eqno(16)$$
and  where the refraction width $d_o$ is given by
$d_o=\sqrt 2\pi G_F^{-1}m_n$.

The reduction in the neutrino flux  when lasing effects are included
can be approximated by
$$\Phi_e(r)^{lase}=\Phi_e^o\exp\biggl [-{1\over 8 d_o}
\sum_x g^2D^{eff} \biggr ] \>,
\eqno(17)$$
where
$$D^{eff}\approx\int^r_0 \chi (r) \rho(r) Y_e(r) dr  \>,\eqno(18)$$
and where
$$\chi (r)= {n_\nu\over EV^2(r)} \>. \eqno(19)$$

We apply Eq.~(17) to a solar model [14], and determine the ratio
${\Phi_e(r)/\Phi_e(r)^{lase}}$.  Our calculations show that  for a
decay such as $\nu_e\rightarrow \nu_x + B$ where the chirality of the
neutrinos remain unchanged, ${\Phi_e(r)/\Phi_e(r)^{lase}}$ can be $>>1$
for both pp and $^8$B neutrinos.  This means that any such decay
applied to the solar neutrinos must take into account this effect.
 For processes  where $\alpha=-1$, for example chirality flipping decay
with spin-zero boson emission, Eq.~(19) becomes
$$\chi (r)= {n_\nu\over E^2V(r)} \>. \eqno(20)$$
 and there is a negligible effect due to the spin suppression effects
discussed earlier.

In order to see the effects of lasing let us assume a
chirality-conserving interaction and apply the current laboratory
bounds on the  chirality-conserving decay  $\nu_e \rightarrow \nu_x+\phi$,
namely $g$ as defined in Eq.~14 is $<7.0\times 10^{-3}$ [15].
 Remembering that the lasing switches off when $f_{\nu_e}=f_{\nu_x}$,
we find that half the pp neutrinos could in fact undergo decay in the
solar interior.   This conclusion remains valid
 for $g \gtorder 10^{-6}$.  Decays of $\nu_e$ to a  left-handed antineutrino
through  majoron emission give a similar result (though a slight
modification to the potential $V$ is required in this case).  Decay
into a right-handed antineutrino would require that the
chirality be flipped, and thus is unimportant due to spin suppression
effects.

A question remains as to whether there is a well-defined model which
couples neutrinos to a boson in such a way that we will see lasing.  As
stated above the most obvious candidate to consider is the singlet
majoron model [16].  Consider a scalar potential of the form
$$
V[\Phi]=\lambda (\Phi^2-v^2)^2\>.\eqno(21)
$$
Clearly the vacuum corresponds to $|\langle\Phi\rangle|=v$, but we have
a number of ways of describing excitations.  If we consider the linear
realization of the symmetry breaking, then we write
$$
\Phi=v+\rho+i\phi \eqno(22)
$$
and find that $\rho$ is a massive scalar field, and that $\phi$ is a
massless Goldstone boson -- the majoron. Including matter effects, we
can determine that part of the lagrangian which involves  the neutrinos
and the linear majoron coupling:
$$
{\cal L}=- \biggl [\bar N_R M_D N_L +{1\over 2} \bar N_R M_R (N_R)^c
+ H.c. \biggr ] +
{i\over 2v} \phi\bar N_R M_R (N_R)^c
-{1 \over \sqrt 2} \zeta_u \bar N_L\gamma^\mu N_L
\>, \eqno(23)$$
where $N_{L(R)}$ are the weak-lepton-doublet (singlet) neutrinos,
$\zeta_u=G_F\rho m_n^{-1} Y_n v_\mu$ (where $v_\mu$ is collective
nucleon four-velocity), and the mass matrices $M_D$ and $M_R$ are
assumed real. Considering only two generations and assuming $M_R>>M_D$,
diagonalization of the mass matrix determines the
physical fields $\nu_{1,2}$.  The coupling of  these neutrino fields to
$\phi$ in this case is direct, and has the form
$$
ig\phi\bar\nu_2 \gamma_5 \nu_1\>,\eqno(24)
$$
meaning that the local operator for the transition $\nu_2\to \nu_1+\phi$
necessarily involves
a chirality flip.  As seen earlier, the spin suppression associated with
this chirality flip renders lasing in the sun unimportant. Only within
environments where the neutrino density is much larger than solar
neutrino densities (eg. supernova explosions), would the coupling of
Eq.~(24) be important.

It is possible to have chirality-conserving decays from other
processes.  In the linear expansion this occurs when mass insertions
are added to the external neutrino lines. However, in the most  simple
models each mass insertion introduces a suppression factor of order
$(m/E)^2$. If the exponential expansion $\Phi=(v+\rho)\exp (i\phi/v)$
is used, the chirality conserving decays naively involve derivative
couplings which suppress the  emission of low momentum bosons, and
hence the lasing rate also.  The actual result must be independent of
which expansion is utilized, but only after all relevant diagrams at
each order are included would the exact suppression factor be
determined. Such suppression would clearly be  model dependent. Indeed,
we note that it is possible to construct  more complicated majoron
models  which  exhibit  no suppression of the  chirality-conserving
decays [18].

Another hopeful possibility is that the neutrino decays through a new
spin--1 field.  The coupling would be between neutrinos of the same
chirality state,  and the rate would not be suppressed by any chirality
or momentum effects.  Massive vector fields were speculated on as
possible sources of lepton number violation through local symmetry
breaking [17], but were superseded by majoron models and global
symmetry breaking, which seemed to arise more naturally in GUT's.

In summary, we have seen how neutrino lasing may play a role in
neutrino decay processes occurring in the solar interior. We have
highlighted how the spin nature of the particles involved in the decay
process play an important role.  In particular, we noted that within
the  context of the most simplest singlet majoron models, no
significant neutrino lasing in the sun would proceed. More complicated
extensions to the standard model need to be invoked if  lasing is to be
viable in the solar interior.
 Since the energy spectrum of the decaying electron neutrinos is not
degraded by the lasing phenomenon, future detectors should be able to
distinguish between the process described here and that anticipated
from the usual decay paradigm.

A similar analysis  of neutrino lasing applied to supernovae explosions
may provide some additional constraints. We note again, however, that
since bosons are preferentially produced at low momentum the energetics
of such explosions should not be significantly altered by excess energy
losses, unless the decay product neutrino is sterile.

Neutrino lasing has the potential to be an important phenomenon.
In addition to its impact on the dark matter problem, it may also have
interesting implications for  majoron  models of baryogenesis due to
enhanced decays of sterile neutrinos. Any experimental information on
the viability of neutrino lasing would therefore be very valuable.

\medskip
\noindent
We thank N. Kaiser, M. Luke, S. Selipsky and L. Widrow for useful discussions.

\vfil
\eject
\noindent{\bf References}

\item{[1]} J. N. Bahcall and M. N. Pinsonneault,
Rev.\ Mod.\ Phys.\ {\bf 64}
(1992) 885, and references therein.

\item{[2]} R.\ Davis,  in Proceedings of the 21$^{\rm st}$
International Cosmic Ray Conference, ed.\ R.J.\ Protheroe
(University of Adelaide press), 1990, vol.\ 12, p.143.

\item{[3]} A. Suzuki, in International Symposium on Neutrino Astrophysics,
KEK preprint 93-96; see also K.S. Hirata {\it et al.},
Phys.\ Rev.\ Lett.\ {\bf 66} (1991) 9.

\item{[4]} V. N. Gavrin, in International Conference on High Energy Physics,
Dallas, Texas, 1992, ed. J. Sanford (AIP, New York 1993).

\item{[5]}  P.\ Anselmann {\it et al}, Phys.\ Lett.\ {\bf B285} (1992) 376.

\item{[6]} J. N. Bahcall, N. Cabibbo and A. Yahill, Phys.\ Rev.\ Lett.\
{\bf 28} (1972) 316.  Other discussions of the decay paradigm can be
found in A.  Acker, A. Joshipura and S. Pakvasa Phys.\ Lett.\ {\bf
B285} (1992) 371; Z. G. Berezhiani and A. Rossi, INFN FE-08-93 preprint
(1993).

\item{[7]}Z. G. Berezhiani {\it et al}, JETP Lett. {\bf 55} (1992) 151

 \item{[8]}Z. G. Berezhiani and M. I. Vysotsky, Phys.\ Lett.\ {B199}
(1987) 281;
J. A. Frieman, H. E. Haber and K. Freese, Phys.\ Lett.\ {\bf B200}
(1988) 115.

\item{[9]} K. Choi and A. Santamaria, Phys.\ Rev.\ {\bf D42} (1990) 293
and references therein.

\item{[10]} A. Acker, J. Panteleone and S. Pakvasa,
Phys.\ Rev.\ {\bf D43} (1991) 1754

\item{[11]} N. Kaiser, R. A. Malaney and G. Starkman, Phys.\ Rev.\ Lett.\
{\bf 71} (1993) 1128.

\item{[12]} G. Starkman,  N. Kaiser and R. A. Malaney, CITA/93/49 preprint,
Ap. J., submitted (1993).

\item{[13]} Z. G. Berezhiani and A. Yu. Smirnov, Phys.\ Lett.\ {\bf B220}
(1989) 279;
for detailed discussion of matter effects see
 L.\ Wolfenstein, Phys.\ Rev.\ D {\bf 17}, (1978) 2369, and
    Phys.\ Rev.\ D {\bf 20}, (1979) 2634;
  S.P.\ Mikheyev and A.Yu.\ Smirnov, Yad.\ Fiz.\ {\bf 42},
      (1985) 1441 [Sov.\ J.\ Nuc.\ Phys.\ {\bf 42}, (1985) 913];
  H.A.\ Bethe, Phys.\ Rev.\ Lett.\ {\bf 56}, (1986) 1305.

\item{[14]} We thank B. Chaboyer and J. Bahcall for supplying us with
standard solar model grids.

\item{[15]}  V. Barger, W. Keung and S. Pakvasa, Phys.\ Rev.\ D {\bf 25}
(1982) 907.

\item{[16]}  Y.\ Chikashige, R.N.\ Mohapatra and R.D.\ Peccei,
Phys.\ Lett.\ {\bf 98B} (1981)
265.

\item{[17]}  E.\ Witten, Phys.\ Lett.\ {\bf 91B}, (1980) 81;
R.N.\ Mohapatra and
G.\ Senjanovi\`c, Phys.\ Rev.\ Lett {\bf 44}, (1980) 912.

\item{[18]} See, for example, Z.G.\ Berezhiani, A.Yu.\ Smirnov and
J.W.F.\ Valle, Phys.\ Lett.\ {\bf B291} (1992) 99.

\vfil
\eject
\bye